# 基于随机效应模型的交叉口事故碰撞类型建模

王雪松[1]，袁景辉[1]，杨筱菡[2]

（1.同济大学 道路与交通工程教育部重点试验室，上海 201804；2.同济大学 数学系，上海 201804）

**摘要**：考虑交叉口不同进口道之间的差异性，在进口道层面对不同碰撞类型事故分别建立随机效应模型，通过在负二项模型中引入随机效应项来考虑交叉口不同进口道之间的空间相关性。模型参数采用全贝叶斯方法进行估计。结果表明：不同碰撞类型事故的影响因素不同，且同一影响因素对不同碰撞类型事故的影响程度也存在显著差异，验证了区分碰撞类型进行事故建模的必要性。随机效应项在各模型中均显著，表明交叉口不同进口道之间事故频率存在显著的空间相关性。

**关键词**：信控交叉口；事故碰撞类型；交叉口进口道；空间相关性；随机效应模型

## Modeling of Crash Types at Signalized Intersections Base on Random Effect Model

*Wang Xuesong[1], Yuan Jinghui[1], Yang Xiaohan[2]*

(1. The Key Laboratory of Road and Traffic Engineering of the Ministry of Education, Tongji University, Shanghai 201804, China; 2. The Department of Mathematics, Tongji University, Shanghai 201804, China)

**Abstract:** Random effect models were developed by crash types at approach-level. A random effect term, which indicates the intersection-specific effect, was incorporated into each crash type model to deal with the spatial correlation between different approaches within the same intersection. The model parameters were estimated under the Bayesian framework. Results show that different crash types are correlated with different groups of factors, and each factor shows diverse effects on different crash types, which indicates the importance of crash type models. Besides, the significance of random effect term confirms the existence of spatial correlations among different approaches within the same intersection.

**Keywords:** signalized intersection; crash types; intersection approach; spatial correlation; random effect model

　　传统的交叉口安全分析通过建立交叉口总事故数与几何设计、信号控制、运行状况等特征变量的统计模型来分析事故发生的显著影响因素。然而不同碰撞类型事故发生前的碰撞行为不同，其影响因素也存在较大差异。例如，追尾事故大多是由于车辆间未保持合理间距以及紧急加减速所导致，而直角侧撞事故大多是由于车辆闯信号灯或者信号配时设置不合理所导致。考虑到针对不同碰撞类型事故多发交叉口需要采取不同的改善措施，基于交叉口总事故数建立安全分析模型难以支撑对不同事故类型多发交叉口分别进行判别，因此有必要区分事故碰撞类型进行建模研究。

　　交叉口各进口道的几何设计、信号控制、交通流量均存在一定差异，且各进口道发生的事故数也并不是均匀分布。在交叉口层面集计各进口道的事故总数与交叉口整体特征建立统计模型容易掩盖事故与进口道层面特征因素的关联关系，难以对事故影响因素进行深入分析。在进行交叉口安全分析时，基于进口道层面能够更好地对事故发生的具体影响因素进行深入研究。

　　同一交叉口的不同进口道的交通流量、信号控制等因素之间存在相互影响，致使不同进口道之间存在一定的空间相关性。传统的建模方法假设样本之间相互独立，会造成不准确的统计推断。本文综合考虑各进口道之间的差异性和相关性，基于交叉口进口道的特征数据对各碰撞类型事故分别进行建模研究。考虑到目前国内只有上海市的事故数据可以定位至交叉口进口道，然而其对于事故碰撞类型的记录不够完善，难以满足区分事故碰撞类型分别建模的要求。为了实现基于事故碰撞类型的交叉口进口道事故建模，本文使用美国佛罗里达州 Orange 县和 Hillsborough 县的数据，为后期国内进行系统性交叉口安全分析提供借鉴。

# 1 研究综述

　　国外围绕交叉口事故建模及其影响因素分析开展了大量的研究，其中大多采用总事故数与交叉口特征数据进行建模分析[1]，未考虑不同碰撞类型事故之间影响因素的差异性。Hauer 等[4]最早开始区分事故碰撞类型进行建模研究，将交叉口事故分为 15 种碰撞类型，采用各类型事故冲突相关的车流量分别对各类型事故进行建模。研究人员主要对直角侧撞事故、正面碰撞事故、追尾事故、侧向刮擦事故、左转事故以及撞压行人事故进行研究[5]。交叉口各进口道的几何设计、信号控制、交通流量和事故数并不完全一致，Hall[5]基于交叉口进口道层面对 14 种事故类型分别建立泊松模型进行安全分析。基于进口道进行建模分析可以更好地分析事故与其影响因素的具体关系，逐渐成为交叉口安全分析的趋势。Wang 等[6]基于进口道层面分别对交叉口直角侧撞事故、左转事故进行建模研究。

　　传统的泊松模型和负二项模型均假设样本相互独立，运用到交叉口进口道层面进行事故建模时并不能考虑各进口道间样本数据的关联性，会造成错误的统计推断。Wang 等[6]对交叉口直角侧撞事故进行建模分析时采用广义估计方程（GEE）来考虑交叉口各进口道样本数据间的相关性。GEE 不同的样本组均采用相同的关联矩阵不能反映不同组别之间的差异。Shankar 等[2]采用随机效应负二项模型进行事故建模分析，通

过在负二项模型的连接函数中引入随机效应项来考虑事故数据的时空相关性。随机效应模型可以有效地处理具有一定相关性的事故数据，近年来逐步运用于交叉口和路段事故建模分析[9]。

国外的研究人员对影响交叉口安全的几何设计、控制特征以及交通状态等因素进行了系统的研究。在几何设计方面，其主要的影响因素有车道数[3]、左转车道偏移[6]、中央分隔带[1]；在交通控制方面，主要的影响因素有左转控制类型[3]、限速值[3]；在交通运行状态方面，主要的影响因素为交通流量[3]。

国内由于事故数据记录信息缺失、交叉口特征数据采集不充分以及建模方法存在局限等因素导致现有的交叉口事故统计模型难以满足对交叉口安全进行深入分析的需求。在事故数据逐步完善的基础上，王雪松等[11]利用广义估计方程考虑同一主干道上交叉口的空间相关性，分析交叉口几何设计、信号控制、运行状况等因素对于安全的影响，首次基于国内交叉口的事故数据以及几何、控制等特征数据进行复杂建模分析。谢琨和王雪松[11]通过建立分层贝叶斯模型来考虑位于同一主干道上交叉口间的关联性，对交叉口和主干道层面的安全影响因素进行分析，并且首次考虑了主干道运行速度特征对交叉口安全的影响。本文基于来自美国的数据，在进口道层面区分事故碰撞类型对交叉口安全影响因素分别进行深入分析，为后期国内对交叉口安全的进一步分析提供借鉴。

# 2 数据准备

从美国佛罗里达州 Orange 县和 Hillsborough 县选取 177 个四枝交叉口，收集各交叉口进口道 2000 年至 2005 年的几何设计、控制属性、交通流量以及各类型事故数据，所选交叉口的几何设计、控制属性在分析年限内均无明显变化。几何设计特征均提取自 Google Earth 高分辨率影像，并据此确定进口道各类型车道的数量、左转车道偏移、相交道路的走向及夹角以及是否有中央分隔带。

通过分析城市交通部门的信号控制方案得到各交叉口进口道的交通控制属性，包括进口道左转控制类型（左转专用、允许冲突、左转可变）、线控类型、黄灯时间、全红时间、进口道限速值、信号灯是否闪烁；两县各交叉口进口道年平均日交通量通过取 2000-2005 年的年平均日交通量的平均值得出，日转向交通量通过将年平均日交通量乘以高峰时段各进口道转向车流比得出。所有自变量在进口道层面的统计性描述如表 1 所示。

表 1 交叉口进口道变量的统计性描述
Tab. 1 Descriptive Statistics of Approach-Level Variables

| 参数 | 变量 | 平均值 | 最小值 | 最大值 | 标准差 |
|---|---|---|---|---|---|
| 运行特征 | 年平均日交通量/辆 | 13042.33 | 51 | 50763 | 10201.57 |
|  | 直行年平均日交通量/辆 | 9280.59 | 10 | 50464 | 9190.20 |
|  | 左转年平均日交通量/辆 | 2075.08 | 10 | 13005 | 2148.97 |
|  | 右转年平均日交通量/辆 | 1686.68 | 3 | 11653 | 1717.87 |
| 几何设计 | 总车道数/条 | 3.32 | 1 | 7 | 1.32 |
|  | 直行车道数/条 | 1.76 | 1 | 5 | 0.81 |
|  | 左转车道数/条 | 1.06 | 0 | 2 | 0.52 |
|  | 右转车道数/条 | 0.50 | 0 | 2 | 0.51 |
|  | 中央分隔带/（0 表示无；1 表示有） | 0.53 | 0 | 1 | 0.50 |
|  | 左转偏移（-1 表示负偏移；0 表示无偏移；1 表示正偏移） | 0.40 | -1 | 1 | 0.69 |
|  | 相交道路夹角/（°） | 90.12 | 36 | 144 | 13.05 |
|  | 路面摩擦值 | 35.68 | 24.19 | 46.07 | 4.01 |
| 交通控制 | 信号线控类型（0 表示单点控制；1 表示线控） | 0.38 | 0 | 1 | 0.49 |
|  | 左转控制类型（0 表示允许冲突；1 表示左转可变；2 表示左转专用） | 0.94 | 0 | 2 | 0.85 |
|  | 黄灯时间/s | 2.46 | 0 | 5.5 | 2.11 |
|  | 全红时间/s | 1.67 | 0.5 | 5.1 | 0.65 |
|  | 信号灯闪烁模式（0 表示无闪烁；1 表示闪烁） | 0.08 | 0 | 1 | 0.27 |
|  | 限速值/（km·h$^{-1}$） | 41.82 | 15 | 60 | 6.95 |
| 行政区域 | 县城（0 表示 Orange 县；1 表示 Hillsborough 县） | 0.64 | 0 | 1 | 0.48 |

注释：路面摩擦值为路面摩擦系数乘以 100 所得；对于信号灯闪烁模式，0 表示信号灯全天无闪烁，1 表示信号灯白天不闪烁，在深夜和凌晨采用闪烁控制。

基于佛罗里达州交通厅事故分析报告系统中检索出的事故记录，提取每个交叉口在 2000 年至 2005 年之间发生的事故。考虑到发生在交叉口停车线上游的事故也会受交叉口影响[13]，根据交叉口的安全影响区范围以及事故发生位置，将 1940 起发生在停车线上游且位于交叉口影响区范围内的事故标记为交叉口事

故，选定的交叉口共有 12318 起事故。本文最终选取 5 种典型的碰撞类型事故进行建模分析，总计 11386 起事故，其中追尾事故（7279 起）、侧向刮擦事故（807 起）、直角侧撞事故（848 起）、对向左转事故（2059 起）、相交左转事故（393 起）。图 1 解释了每种碰撞类型事故的冲突模式。

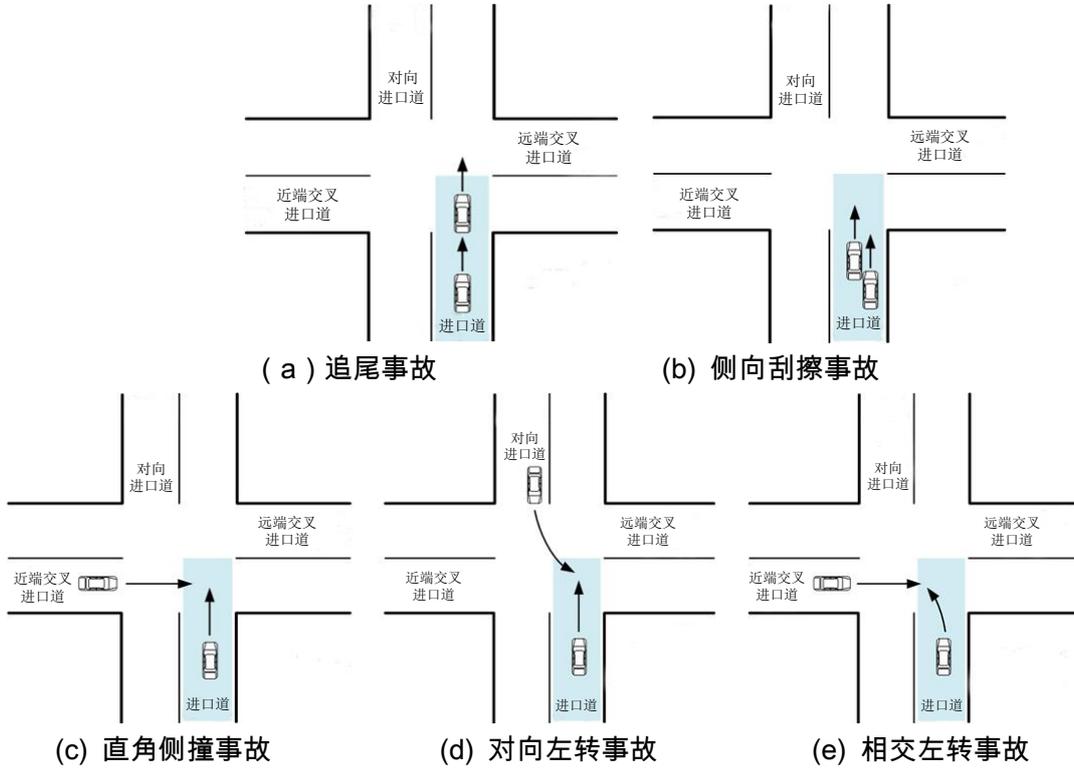

（a）追尾事故　　（b）侧向刮擦事故

(c) 直角侧撞事故　　(d) 对向左转事故　　(e) 相交左转事故

图 1 各碰撞类型事故的车辆冲突模式

Fig. 1 Crash Types Classified by Conflicting Vehicle Maneuvers

## 3 基于全贝叶斯方法的随机效应模型

本研究采用负二项模型来处理离散的事故数据，设 $y_{ij}$ 为 $i$ 交叉口 $j$ 进口道的事故数，负二项模型中 $y_{ij}$ 服从负二项分布，如下所示：

$$y_{ij} \sim \text{Negbin}(\theta_{ij}, r) \tag{1}$$

式中：$\theta_{ij}$ 为 $y_{ij}$ 的期望；$r$ 为离散系数。

连接函数为对数函数，原始的负二项模型如下所示：

$$\log(\theta_{ij}) = \boldsymbol{\beta}\boldsymbol{X}_{ij} \tag{2}$$

式中：$\boldsymbol{X}_{ij}$ 为交叉口进口道层面的解释变量的向量；$\boldsymbol{\beta}$ 为回归系数的向量。

考虑到同一交叉口各进口道由于信号控制以及交通流量的相互影响而导致事故观测值存在空间相关性，在负二项模型中引入一个随机效应项 $\varphi_i$，以解释 $i$ 交叉口内不同进口道的空间相关性。随机效应模型表达式如下所示：

$$\log(\theta_{ij}) = \boldsymbol{\beta}\boldsymbol{X}_{ij} + \varphi_i \tag{3}$$

$$\varphi_i \sim \text{N}(0, \sigma_\varphi^2) \tag{4}$$

式中：$\sigma_\varphi^2$ 为随机效应 $\varphi_i$ 服从正态分布的方差。

全贝叶斯方法结合先验分布以及从观测数据得出的似然函数，得出估计参数的后验分布[14]。先验分布可以根据经验给出，也可以是无信息的先验分布。

全贝叶斯方法的理论框架可以表示为

$$\pi(\boldsymbol{\theta} | \boldsymbol{y}) = \frac{L(\boldsymbol{y} | \boldsymbol{\theta})\pi(\boldsymbol{\theta})}{\int L(\boldsymbol{y} | \boldsymbol{\theta})\pi(\boldsymbol{\theta})\mathrm{d}\boldsymbol{\theta}} \tag{5}$$

式中：$\pi(\boldsymbol{\theta} | \boldsymbol{y})$ 为给定 $\boldsymbol{y}$ 条件下 $\boldsymbol{\theta}$ 的后验分布，$\boldsymbol{y}$ 为样本观测数据的向量，$\boldsymbol{\theta}$ 是似然函数系数的向量；$L(\boldsymbol{y}|\boldsymbol{\theta})$ 为似然函数；$\pi(\boldsymbol{\theta})$ 为 $\boldsymbol{\theta}$ 的先验分布；$\int L(\boldsymbol{y}|\boldsymbol{\theta})\pi(\boldsymbol{\theta})\mathrm{d}\boldsymbol{\theta}$ 为观测数据的边缘概率分布。

本文采用全贝叶斯方法对随机效应模型参数进行估计，全贝叶斯方法通过引入估计参数的先验分布来避免由于样本数据随机性导致的回归平均值问题。由于没有可靠的先验信息，假定所有的回归系数服从正态分布 $N(0,10^5)$，负二项分布离散系数 $r$ 以及随机效应的方差 $\sigma_\varphi^2$ 服从 Inverse-Gamma 分布 $(10^{-3},10^{-3})$。

# 4 基于碰撞类型的交叉口进口道事故建模

考虑到不同碰撞类型事故发生频率与冲突交通量有关[4]。本文中各碰撞类型模型变量中交通流量均采用各碰撞类型的冲突交通量，具体如表 2 所示。

贝叶斯估计常通过马尔科夫链算法(MCMC)[15]来完成，吉布斯取样法被广泛的应用于马尔科夫链的模拟[16]。采用 WinBUGS 软件基于吉布斯取样法来完成贝叶斯模型的标定，考虑到系数收敛和迭代时间，设定两条马尔科夫链进行 20000 次迭代，舍弃前 2000 个不稳定的样本，各碰撞类型事故的随机效应模型参数标定的后验分布结果如

表 3 所示，表中已剔除 95% 置信区间包含零的不显著变量。

表 2 各碰撞类型模型中的交通流量组成
Tab.2 Traffic Volume Types for Each Crash Type Model

| 碰撞类型事故模型 | 交通流量 |
| --- | --- |
| 追尾事故 | 进口道交通量 |
| 对向左转事故 | 进口直行和对向左转交通量之积 |
| 相交左转事故 | 进口左转和近端交叉直行交通量之积 |
| 直角侧撞事故 | 进口直行与近端交叉直行交通量之积 |
| 侧向刮擦事故 | 进口道交通量 |

表 3 各碰撞类型模型参数的后验分布
Table.3 Posterior Summary of Each Crash Type Model

| 变量 | | 追尾事故 | 对向左转事故 | 相交左转事故 | 直角侧撞事故 | 侧向刮擦事故 |
| --- | --- | --- | --- | --- | --- | --- |
| 常数项 | | -4.51（0.4755） | -5.99（0.7025） | -6.88（0.8952） | -2.416（0.4293） | -7.535（0.7538） |
| 对数交通量 | | 0.658（0.0436） | 0.2829（0.0457） | 0.3828（0.0463） | 0.1748（0.0266） | 0.6466（0.0906） |
| 进口道左转车道数 | | - | - | - | - | 0.3842（0.1158） |
| 进口道直行车道数 | | - | 0.1906（0.0853） | - | -0.1（0.0557） | 0.1975（0.0874） |
| 进口道右转车道数 | | 0.2522（0.0627） | - | - | - | 0.2615（0.1025） |
| 对向进口道直行车道数 | | - | - | -0.5972（0.1518） | - | - |
| 进口道有无中央分隔带 | | - | 0.4062（0.1401） | -0.1748（0.2041） | - | - |
| 进口道路面摩擦值 | | -0.0127（0.0104） | - | - | - | - |
| 信号线控类型 | | 0.2394（0.0738） | - | - | - | - |
| 进口道左转控制类型 | 允许冲突（基准） | | | | | |
| | 左转专用 | 0.681（0.1014） | -0.5272（0.1716） | 0.2788（0.2368） | - | 0.5328（0.1728） |
| | 左转可变 | 0.3728（0.0912） | 0.4506（0.1458） | 0.2033（0.1953） | - | 0.4613（0.1537） |
| 黄灯时间与标准值之差 | | - | - | - | -0.3945（0.1287） | - |
| 全红时间与标准值之差 | | - | - | - | -0.1137（0.0754） | - |
| 信号灯闪烁模式 | | - | - | - | 0.5（0.1999） | - |
| 进口道限速值 | | 0.01（0.0053） | 0.0434（0.0087） | - | - | - |
| 近端交叉进口道限速值 | | - | - | 0.0228（0.0128） | - | - |
| 离散系数 | | 0.2319（0.0257） | 0.7083（0.0851） | 0.5690（0.1934） | 0.1238（0.0657） | 0.064（0.0518） |

| 随机效应的方差 | 0.2816（0.0464） | 0.4600（0.1020） | 0.3341（0.1408） | 0.2908（0.0681） | 0.3677（0.0879） |

注释：括号内数字为模型参数估计的标准差；对数交通量中的交通量为各碰撞类型对应的冲突相关交通量，具体如表 2 所示。

## 4.1 安全影响因素分析

对比各碰撞类型模型的参数估计结果可知，不同碰撞类型事故的显著影响因素均不相同，同一影响因素对不同碰撞类型事故的影响程度也存在较大差异。例如左转专用控制对追尾事故、相交左转事故、侧向刮擦事故的影响均显著为正，而对对向左转事故的影响显著为负。模型结果进一步验证了区分碰撞类型进行事故建模的必要性。

总体而言，交叉口安全的显著影响因素主要有三个方面：交通特征、几何设计、交通控制，以下分别对各类影响因素进行分析。

### 4.1.1 交通特征

冲突交通流量对各碰撞类型事故的影响均显著为正，这与以往的研究结论一致[4]。Wang 等[6]在对直角侧撞事故、左转事故分别进行的研究中均得出了相同的结论。建模结果表明冲突交通量对追尾事故（0.658）和侧向刮擦事故（0.6466）的影响程度较其他三类事故要高（括号内为变量对某碰撞类型事故的影响系数，以下类同），原因可能是追尾事故与侧向刮擦事故均为进口道内车辆间碰撞，冲突交通量在时空上未分离，发生事故受交通流影响较大；而另外三类事故的冲突模式均为不同进口道的车辆间碰撞，在理想的信号控制情况下，冲突交通在时间上分离，因此冲突交通量对这三类事故发生的影响程度较低。

### 4.1.2 几何设计

右转车道数仅对追尾事故和侧向刮擦事故有显著影响。右转车道数与追尾事故显著正相关，这与以往的研究类似[8]。原因可能是由于右转车道不受信号控制，需要避让其他方向车辆以及慢行交通，刹车频率较高，容易导致追尾事故发生；右转车道数对侧向刮擦事故数有显著的正影响。进口道的车道数越多，断面行驶车辆越多，不同车道间的车辆潜在接触面越多，容易引发侧向刮擦事故。

左转车道数则只对侧向刮擦事故有显著正影响。左转车辆越多，则进口道内的车辆变道行为越频繁，容易引发侧向刮擦事故。Wang 等[8]在研究交叉口追尾事故时，发现左转车道数对追尾事故也有显著影响。

直行车道数对直角侧撞事故、侧向刮擦事故以及对向左转事故均有显著影响。直行车道数与直角侧撞事故数显著负相关，随着进口道直行车道数的增加，近端交叉直行车辆穿过交叉口所需行驶距离更长，发生冲突的几率更大，然而对于近端交叉进口道直行驾驶员而言，考虑到由于距离较长而带来的危险性增加，驾驶员可能会倾向于保守驾驶而不去闯红灯，从而直角侧撞事故会有一定的减少；直行车道数对侧向刮擦事故有显著正影响，原因与右转车道数类似；直行车道数与对向左转事故数显著正相关，进口道直行车道越多，意味着直行车辆流量越大，因此其与对向左转车辆的冲突概率越大[7]；

对向直行车道数仅对相交左转事故有显著负影响。原因可能是对向进口道直行车道数越多，进口道车辆左转所需跨越的距离越长，考虑到暴露在交叉口冲突区域的距离较长，左转车辆驾驶员更偏向于保守驾驶，因此相交左转事故频率越低。

有无中央分隔带仅对相交左转事故和对向左转事故有显著影响。有中央分隔带的进口道发生相交左转事故的可能性较低，车辆左转所需跨越距离较长，因此左转驾驶员会更倾向于遵守交通规则；有中央分隔带的进口道更容易发生对向左转事故。由于中央分隔带的存在，对向左转车辆对于进口道直行车辆的视距比没有中央分隔带的进口道要差，会增加车辆左转的危险性。

### 4.1.3 控制属性

进口道左转控制类型对追尾事故、对向左转事故、相交左转事故以及侧向刮擦事故均有显著影响。对追尾事故而言，左转专用控制（0.681）比左转允许冲突控制增加了一个左转保护相位，随着周期内相位的增加，追尾事故发生的概率会显著提高[8]。类似地，采用左转可变控制（0.3728）的进口道发生追尾事故的概率同样比左转允许冲突的进口道高。

对于对向左转事故，左转专用控制（-0.5272）的进口道在时间上将进口道直行车流与对向进口道左转车流进行分离，很好的避免了相互之间的冲突，从而会降低对向左转事故发生的概率；而左转可变控制（0.4506）比单独的左转允许冲突控制或左转专用控制都要复杂，容易造成进口道直行车辆驾驶员的违规以及对向进口道左转车辆驾驶员的错误判断，从而增加发生事故的风险[7]。

对于相交左转事故，随着进口道左转保护等级的提高，近端交叉进口道的直行车辆违规的比例越高，反而更容易引发相交左转事故；对于侧向刮擦事故，随着左转保护等级的提高，周期内相位数增多，进口道内车辆停车次数增加，且直行车辆与左转车辆的通行时间有错开，更容易引发侧向刮擦事故。

信号线控类型仅对追尾事故有显著正影响。采用信号线控的交叉口由于对上下游的交叉口信号灯进行

了协调控制，因此行驶车辆需要等待的红灯次数更少，车辆行驶速度越高，从而更容易引发追尾事故[8]。

信号灯闪烁模式仅对直角侧撞事故有显著正影响。采用信号灯闪烁控制的进口道，可能会存在车辆在绿灯闪烁时加速通过交叉口与近端交叉直行车流产生冲突，或者在红灯开始闪烁时便驶入交叉口与近端交叉直行车流产生冲突，从而增加发生直角侧撞事故的概率[6]。

黄灯时长以及全红时长均只对直角侧撞事故有显著负影响。随着黄灯时间增加，每个周期交叉口内部进口道直行车辆与近端交叉进口道直行车辆的冲突车流明显减少，发生直角侧撞事故的概率越低。类似地，随着全红时间加长，交叉口内部的各向冲突车流明显减少，降低发生直角侧撞事故的几率[6]。

进口道限速值仅对追尾事故和对向左转事故有显著正影响。进口道限速值越高，一般意味着进口道的断面平均车速越高。在高速条件下，普通的跟车距离不足以应对随时发生的紧急情况，容易引发追尾事故；类似地，随着进口道限速值的提高，进口道直行车辆平均行驶速度越高，未受左转保护的对向进口道左转车辆所能通行的间隙越少，直行车辆与对向左转车辆的发生冲突的概率越大。

## 5 结语

本文基于交叉口进口道层面针对 5 种碰撞类型事故分别建立了随机效应模型，结果表明不同碰撞类型事故的影响因素不同，同一影响因素对不同的碰撞类型事故的影响程度均存在显著差异，甚至出现截然相反的情况。这也从结论的角度证实了对交叉口区分事故碰撞类型进行建模分析的重要性以及必要性。

相较以往基于交叉口层面的研究，基于进口道层面对各类型事故的影响因素进行分析可以更深入地揭示交叉口进口道的特征属性对各类型事故发生的影响。在此基础上，可以为相关部门的多方面工作提供理论依据，例如交叉口设计、管理以及事故多发交叉口判别与改善分析等。在交叉口设计和管理方面，对车道数、中央分隔带的设计以及信号控制方案的设置将充分考虑其对安全的影响；在事故多发交叉口判别与治理方面，基于进口道的数据运用该模型预测各进口道的各类型事故数，并基于预测结果进行黑点判别，鉴别出对应于不同碰撞类型的事故多发进口道，从而能够更好地针对不同类型事故黑点实施差异化的安全改善措施。

**参考文献**:

**References:**
[1] Chin H C, Quddus M A. Applying the random effect negative binomial model to examine traffic accident occurrence at signalized intersections [J]. Accident Analysis and Prevention, 2003, 35(2): 253.
[2] Shankar V N, Albin R B, MILTON J C, et al. Evaluating median cross-over likelihoods with clustered accident counts: an empirical inquiry using random effects negative binomial model [J]. Transportation Research Record, 1998, 1635: 44.
[3] Abdel-Aty M., Wang X. Crash estimation at signalized intersections along corridors analyzing spatial effect and identifying significant factors [J] .Transportation Research Record,2006,1953:98.
[4] Hauer E, Ng J C N, Lovell J. Estimation of safety at signalized intersections [J]. Transportation Research Record, 1988, 1185: 48.
[5] Hall R. Accidents at Four-arm Single Carriageway Urban Traffic Signals [R]. Wokingham: Transport Research Laboratory, 1986.
[6] Wang X, Abdel-Aty M. Right-angle crash occurrence at signalized intersections [J]. Transportation Research Record, 2007, 2019:156.
[7] Wang X, Abdel-Aty M. Modeling left-turn crash occurrence at signalized intersections by conflicting pattern [J]. Accident Analysis and Prevention, 2008, 40(1): 76.
[8] Wang X, Abdel-Aty M. Temporal and spatial analyses of rear-end crashes at signalized intersections [J]. Accident Analysis and Prevention, 2006, 38(6): 1137.
[9] Xie K, Wang X, Huang H, et al. Corridor-level signalized intersection safety analysis in shanghai, china using bayesian hierarchical models [J]. Accident Analysis and Prevention, 2013, 50: 25.
[10] Wang X, Song Y, Schultz G. Safety analysis of suburban arterials in shanghai [J]. Accident Analysis and Prevention. 2014, 70: 215.
[11] 王雪松，谢琨，陈小鸿等.考虑空间相关性的信控交叉口安全分析[J]. 同济大学学报：自然科学版, 2012, 40(12) : 1814.
WANG Xuesong, XIE Kun. CHEN Xiaohong, et al. Risk factor analysis for signalized intersection along corridors with a consideration of spatial correlation [J]. Journal of Tongji University: Natural Science, 2012, 40(12): 1814.
[12] 谢琨，王雪松.基于分层贝叶斯模型的信控交叉口安全分析[J].中国公路学报, 2014, 27(2): 90.
XIE Kun，WANG Xuesong. Signalized intersection safety analysis using bayesian hierarchical model [J].China Journal of Highway and Transport, 2014, 27(2): 90.
[13] Wang X, Abdel-Aty M, Nevarez A. Investigation of safety influence area for four-legged signalized intersections: nationwide survey and empirical inquiry [J]. Transportation Research Record, 2008, 2083: 86.
[14] Carlin B P, Louis T A. Bayesian methods for data analysis [M]. Boca Raton: Chapman & Hall/CRC, 2009.
[15] Gilks W R, Richardson S, Spiegelhalter D J. Markov chain monte carlo methods in practice [M]. New York: Chapman & Hall, 1995.
[16] Spiegelhalter D J , Thomas A , BEST N G ,et al. WinBUGS version 1.4.1 user manual [M], Cambridge：MRC Biostatistics Unit,

2003.